\documentclass[twocolumn,tighten,times]{aastex62}

\newcommand{\HII}{H {\small{II}} }
\newcommand{\kms}{{\rm km~s}^{-1}}
\newcommand{\Msun} {M_{\sun}}

\received{xx xx, 2019}
\revised{xx xx, 2019}
\accepted{xx xx, 2019}

\begin{document}

\title{First embedded cluster formation in California molecular cloud}

\author{Jin-Long Xu}
\email{xujl@bao.ac.cn, pjiang@bao.ac.cn}
\affil{National Astronomical Observatories, Chinese Academy of Sciences, Beijing 100101, China}
\affil{CAS Key Laboratory of FAST, National Astronomical Observatories, Chinese Academy of Sciences, Beijing 100101, China}

\author{Ye Xu}
\affiliation{Purple Mountain Observatory, Chinese Academy of Sciences, Nanjing 210008, China}

\author{Peng Jiang}
\affiliation{National Astronomical Observatories, Chinese Academy of Sciences, Beijing 100101, China}
\affil{CAS Key Laboratory of FAST, National Astronomical Observatories, Chinese Academy of Sciences, Beijing 100101, China}

\author{Ming Zhu}
\affiliation{National Astronomical Observatories, Chinese Academy of Sciences, Beijing 100101, China}
\affil{CAS Key Laboratory of FAST, National Astronomical Observatories, Chinese Academy of Sciences, Beijing 100101, China}

\author{Xin Guan}
\affiliation{National Astronomical Observatories, Chinese Academy of Sciences, Beijing 100101, China}
\affil{CAS Key Laboratory of FAST, National Astronomical Observatories, Chinese Academy of Sciences, Beijing 100101, China}

\author{Naiping Yu}
\affiliation{National Astronomical Observatories, Chinese Academy of Sciences, Beijing 100101, China}

\author{Guo-Yin Zhang}
\affiliation{National Astronomical Observatories, Chinese Academy of Sciences, Beijing 100101, China}

\author{Deng-Rong Lu}
\affiliation{Purple Mountain Observatory, Chinese Academy of Sciences, Nanjing 210008, China}

\begin{abstract}
We performed a multi-wavelength observation toward LkH$\alpha$ 101 embedded cluster and its adjacent  85$^{\prime}\times$ 60$^{\prime}$ region. The LkH$\alpha$ 101 embedded cluster is the first and only one significant cluster in California molecular cloud (CMC). These observations have revealed that the LkH$\alpha$ 101 embedded cluster is just located at the  projected intersectional region of two filaments. One filament is the highest-density section of the CMC,  the other is a new identified filament with a low-density gas emission. Toward the  projected intersection, we find the bridging features connecting the two filaments in velocity, and identify a V-shape gas structure. These agree with the scenario that the two filaments are colliding with each other. Using the Five-hundred-meter
Aperture Spherical radio Telescope (FAST), we measured that the RRL velocity of the LkHα 101 H II region
is 0.5 $\kms$, which is related to the velocity component of the CMC filament. Moreover, there are some YSOs distributed outside the intersectional region. We suggest that the cloud-cloud collision together with the fragmentation of the main filament may play an important role in the YSOs formation of the cluster.
\end{abstract}

 \keywords{ISM: clouds -- stars: formation -- ISM: kinematics and dynamics -- ISM: individual objects (LkH$\alpha$ 101)}

\section{Introduction} \label{sec:intro}
High-mass stars ($>$8 $\Msun$) play an important role in promoting the evolution of their host galaxies 
\citep{Zinnecker2007}. However, how the high-mass stars form is not yet well understood.  Most high-mass stars usually form in stellar clusters \citep{Lada2003}. In order to understand how a high-mass star forms, we must understand the formation of its host embedded cluster. Recently, young embedded cluster-forming systems are often found to be associated with the hub-like multiple filamentary structure. Hence, the filament-hub accretion can be used to explain the formation of the young embedded cluster \citep{Myers2009,Kirk2013}. In addition,  individual embedded clusters often present small age spreads \citep{Jeffries2011,Getman2014}, indicating that a whole cluster of stars precipitate the quasi-instantaneous formation \citep{Elmegreen2000,Tan2006}. This inquires that a significant amount of dense gas need to be rapidly gathered in the place where the cluster was born. Collisions between molecular clouds and external effects (e.g., \HII region and supernovae) can satisfy the condition for rapid gas accumulation \citep{Elmegreen1998, Inoue2013}. The cloud-cloud collisions have been  invoked to explain the formation of super star clusters  \citep{Furukawa2009,Fukui2014,Fukui2016}. 

California molecular cloud (CMC) presents a filamentary structure  extended over about 10 deg in the plane of sky \citep{Lada2009}. In the Perseus constellation, the CMC is considered as the most massive giant molecular cloud in the range of 0.5 kpc from the Sun \citep{Lada2009}. Despite its large size and mass, the CMC appears to be very modest in its star formation  activity. In the CMC, the LkH$\alpha$ 101 embedded cluster is the first and only one significant cluster, which is also consistent with the LkH$\alpha$ 101 \HII region \citep{Barsony1990,Dzib2018}. In the embedded cluster, the star LkH$\alpha$ 101 is a young high-mass star with a spectral type of near B0.5 \citep{Herbig2004,Wolk2010}. Compared to the Orion nebula containing many OB stars, \citet{Lada2017} suggested that the CMC is a sleeping giant. Hence, the CMC provides us an ideal laboratory for investigating the formation of embedded cluster.  In this Letter, we performed a multi-wavelength study to investigate the formation imprint of the LkH$\alpha$ 101 embedded cluster in the CMC. 

\section{Observation and data processing}
To show the molecular gas distribution surrounding the LkH$\alpha$ 101 embedded cluster, we mapped a  85$^{\prime}\times$ 60$^{\prime}$ region centered at position of the LkH$\alpha$ star in the transitions of $^{12}$CO $J$=1-0, $^{13}$CO $J$=1-0 and C$^{18}$O $J$=1-0 lines using the Purple Mountain Observation (PMO) 13.7 m radio telescope, during May and December 2019. The 3$\times$3 beam array receiver system in single-sideband (SSB) mode was used as front end. The back end is a fast Fourier transform spectrometer (FFTS) of 16384 channels with a bandwidth of 1 GHz, corresponding to a velocity resolution of 0.16 km s$^{-1}$ for $^{12}$CO $J$=1-0, and 0.17 km s$^{-1}$ for $^{13}$CO $J$=1-0 and C$^{18}$O $J$=1-0. $^{12}$CO $J$=1-0 was observed at the upper sideband with a system noise temperature (Tsys) of $\sim$280 K, while $^{13}$CO $J$=1-0 and C$^{18}$O $J$=1-0 were observed simultaneously at the lower sideband with a  system noise temperature of $\sim$150 K.  The half-power beam width (HPBW) was 53$^{\prime\prime}$ at 115 GHz. The pointing accuracy of the telescope was better than 5$^{\prime\prime}$. Mapping observations use the on-the-fly mode with a constant integration time of 14 seconds at each point and with a $0.5^{\prime}\times0.5^{\prime}$ grid. The reference-point position is at l= 165.2993$^{\circ}$ and b= -9.2197$^{\circ}$. The standard chopper-wheel method was used to calibrate
the antenna temperature \citep{Ulich1976}. The calibration errors are estimated to be within 10\%.

Moreover, we observed the LkH$\alpha$ 101 \HII region associated with the LkH$\alpha$ 101 embedded cluster in the C168$\alpha$ (1375.28630 MHz) and H168$\alpha$ (1374.60043 MHz) radio recombination lines (RRLs) using the Five-hundred-meter Aperture Spherical radio Telescope (FAST), during  September 2019. FAST is located in Guizhou, China. The aperture of the telescope is 500 m and the effective aperture is about 300 m.  The half-power beam width (HPBW) was 2.9$^{\prime}$ and the velocity resolution  is 0.1 km s$^{-1}$ for the digital backend at 1.4 GHz. Single pointing observation of the LkH$\alpha$ 101 \HII region was taken with the narrow mode of spectral backend. This mode has 65536 channels in 31.25 MHz bandwidth. During observations, system temperature was around 18 K. The observed position was integrated for 60 minutes. Reduced root mean square (rms) of the final spectrum is about 11.0 mK. \citet{Jiang2019} gave more details of the FAST instrumentation.

To trace the ionized-gas distribution of the LkH$\alpha$ 101 \HII region, we  also use the 1.4 GHz radio continuum emission data, obtained from the NRAO VLA Sky Survey (NVSS) with a noise of about 0.45 mJy/beam and a resolution of 45$^{\prime\prime}$ \citep{Condon1998}. In order to trace the polycyclic aromatic hydrocarbon (PAH) emission,  we also utilized the 12 $\mu$m (W3) infrared data from the survey of the Wide-field Infrared Survey Explorer \citep[WISE;][]{Wright2010}. The infrared band 12 $\mu$m has the angular resolutions of 6$\farcs5$.

\begin{figure}
\centering
\includegraphics[width = 0.40 \textwidth]{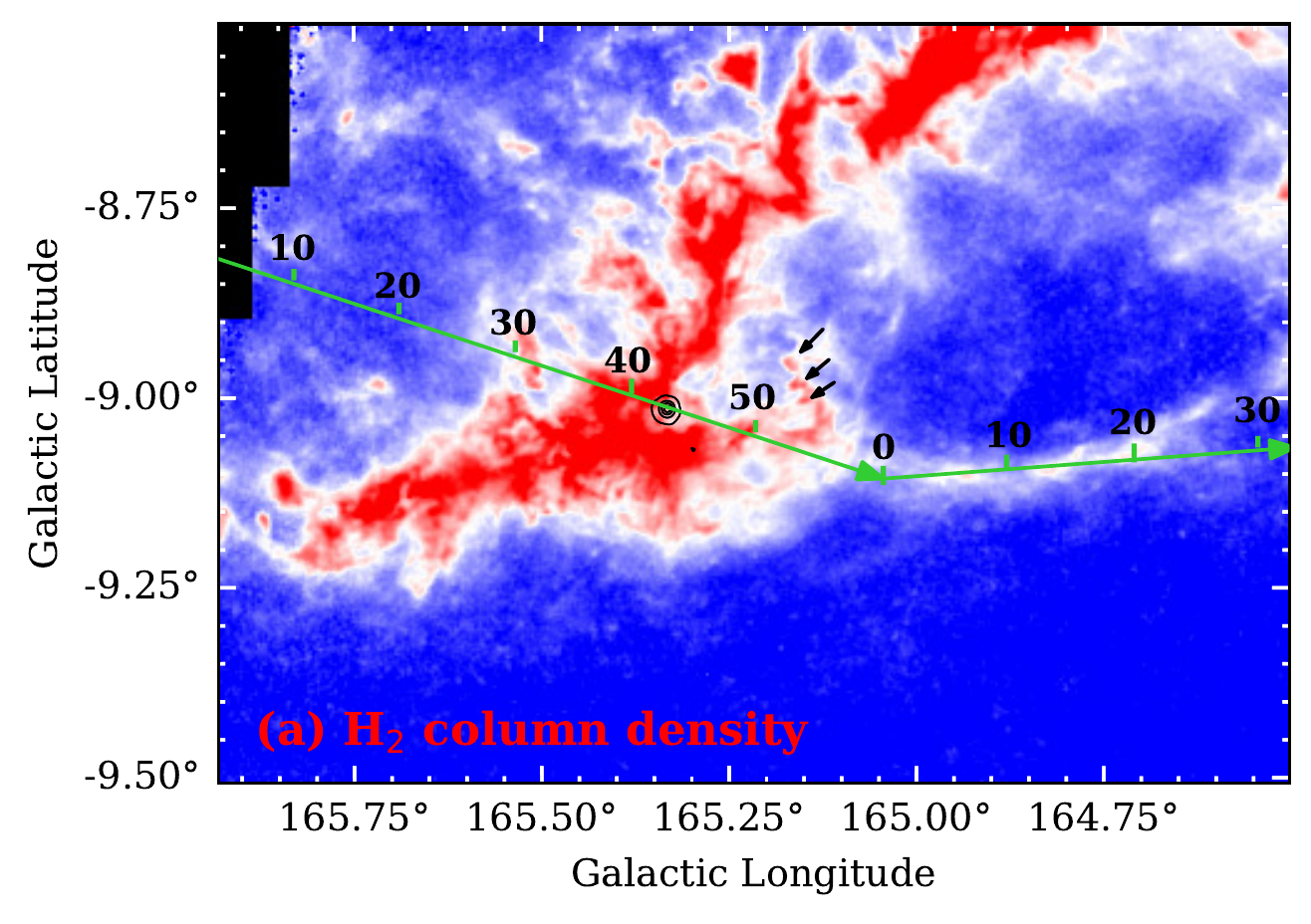}
\includegraphics[width = 0.40 \textwidth]{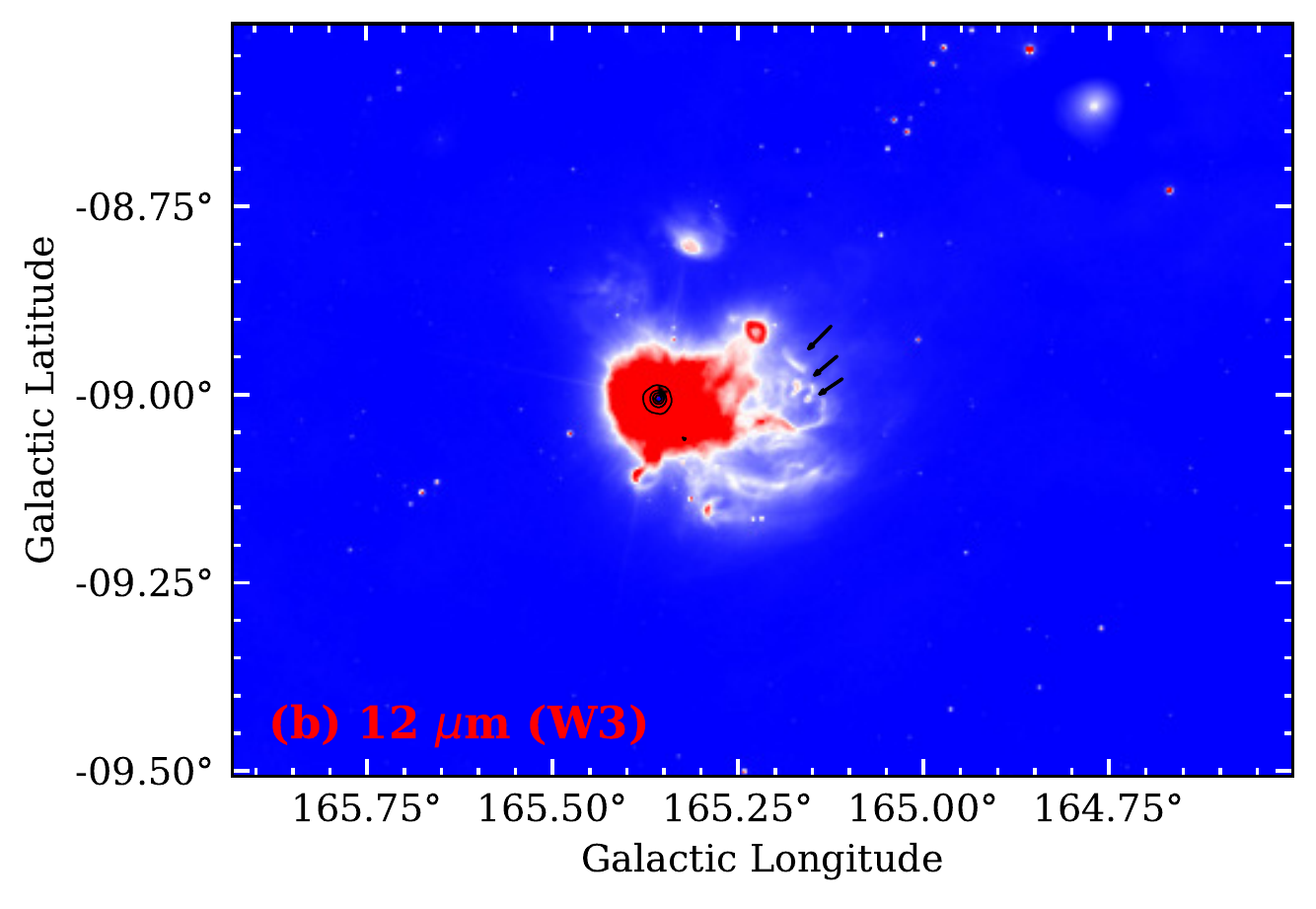}
\vspace{-4mm}
\caption{1.4 GHz radio continuum contours in black colour, overlaid on the Herschel H$_{2}$ column density and WISE 12 $\mu$m images of the observed region in (a) and (b) panels, respectively. The black contours begin at 5$\sigma$ in steps of 10$\sigma$, with 1$\sigma$ = 0.7 mJy beam$^{-1}$.  The  black arrows mark the positions of several pillars. The two green arrows mark the direction and position of position-velocity diagram in Figure \ref{Fig:PV}. We also label the offsets (y-axis in Figure \ref{Fig:PV}) over the two arrows.}
\label{Fig:SH2-222-RGB}
\end{figure}

\section{Results}
\label{sect:results}
\subsection{Infrared and Radio Continuum Images}
Figure \ref{Fig:SH2-222-RGB}(a) shows an H$_{2}$ column density map for our observed region. 
\citet{Zhang2019} constructed the  H$_{2}$ column density map with a resolution of 18.2$^{\prime\prime}$ by using the {\it Herschel} data toward the entire CMC.  Here, we only cut out a 85$^{\prime}\times$ 60$^{\prime}$ region centered at position of the LkH$\alpha$ 101 star from the H$_{2}$ column density map. In Figure \ref{Fig:SH2-222-RGB}(a), the H$_{2}$ column density map shows a filamentary structure elongated from southeast to northwest, named main filament. The main filament is consistent with a filamentary dark cloud L1482  \citep{Harvey2013,Li2014}, which is the highest-density part of the CMC \citep{Lada2017,Zhang2019}. As marked by the two green long arrows in Figure \ref{Fig:SH2-222-RGB}(a), we identify a new filament with a low-density gas emission, named minor filament. The minor filament appear to be perpendicular with the main filament in the line of sight.

\begin{figure}
\centering
\includegraphics[angle=0, width = 0.40 \textwidth]{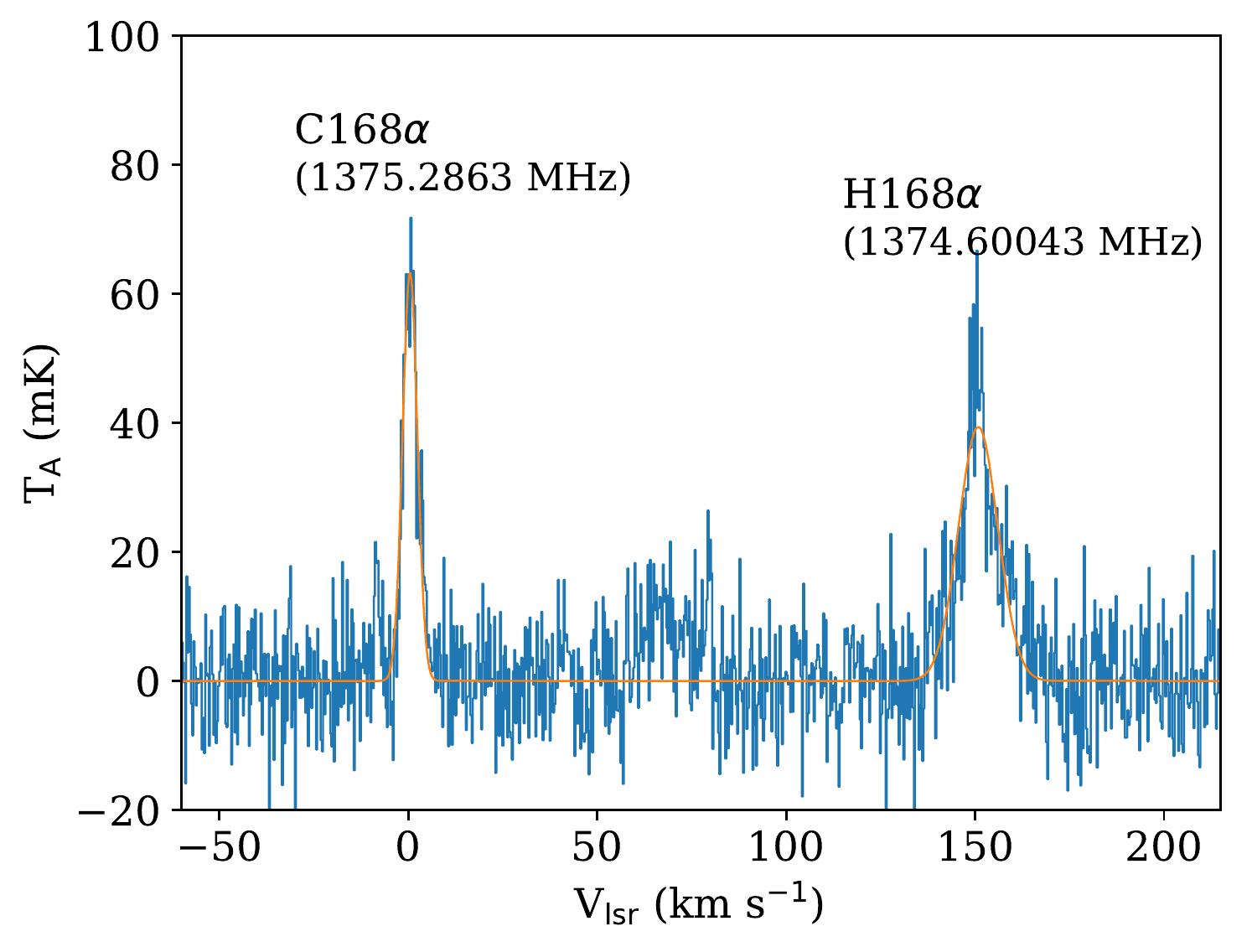}
\vspace{-4mm}
\caption{The C168$\alpha$ and H168$\alpha$ RRLs spectra of the LkH$\alpha$ 101 \HII region from the FAST observations. The red smooth curves is a double-Gaussian fit to the observed spectra. The velocity resolution is smoothed to 0.2 km s$^{-1}$, then the noise RMS is  7.8 mk.}
\label{Fig:RRL}
\end{figure}

The 1.4 GHz radio continuum emission can be used to trace ionized gas. In Figure \ref{Fig:SH2-222-RGB}(a), the ionized-gas emission in black contours of the LkH$\alpha$ 101 \HII region shows a compact structure, which is located at the intersectional region of the main and minor filaments. The WISE 12 $\mu$m contains PAH emission \citep{Tielens2008}.  The PAH molecules are excited by the UV radiation from the \HII region.  In Figure \ref{Fig:SH2-222-RGB}(b), the PAH emission shows a bubble-like structure and several pillars, indicating that the expanding bubble created by the LkH$\alpha$ 101 \HII region  is interacting with the main filament. Figure \ref{Fig:RRL} shows the C168$\alpha$ and H168$\alpha$ RRLs spectra of the LkH$\alpha$ 101 \HII region from the FAST observations.  We use the double-Gaussians to fit the data. By comparing with the H168$\alpha$ RRL, the C168$\alpha$ RRL with a narrow line width is more suitable for tracing the RRL velocity of ionized gas. Hence, we obtain the RRL velocity of 0.5$\pm$0.1 $\kms$ for the LkH$\alpha$ 101 \HII region. 

\subsection{Molecular Gas Images}
\begin{figure}
\centering
\includegraphics[angle=-90, width = 0.25 \textwidth]{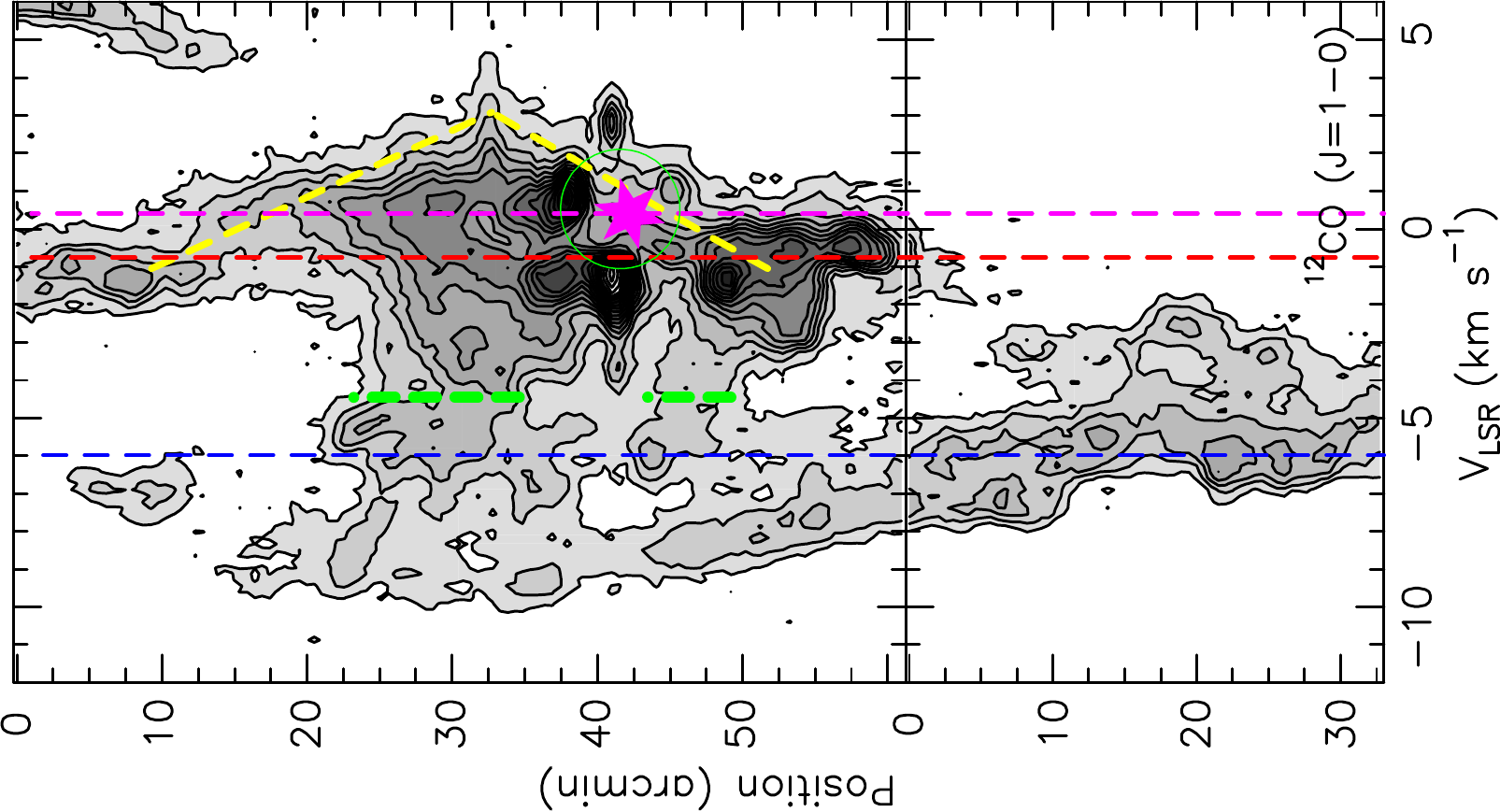}
\caption{Position-velocity diagram of the $^{12}$CO $J$=1-0 emission along the minor filament, as shown in the black dashed line in Figure \ref{Fig:SH2-222-RGB}a. The red dashed line marks the systemic velocity of -0.7 km s$^{-1}$ for the main filament. The purple dashed line indicates the RRL velocity of 0.5 km s$^{-1}$ for the LkH$\alpha$ 101 \HII region. The blue dashed line indicates the systemic velocity of -6.0 km s$^{-1}$ for the minor filament. The contour levels are 10, 20,..., $90\%$ of the peak value. The bridging  positions are indicated by green dashed lines. The purple star represents the LkH$\alpha$ 101 \HII region  or LkH$\alpha$ 101 embedded cluster. The green circle indicate a molecular shell. A ``V-shape'' gas structure is indicated by the yellow dashed lines.}
\label{Fig:PV}
\end{figure}

\begin{figure*}
\centering
\includegraphics[width = 0.65 \textwidth]{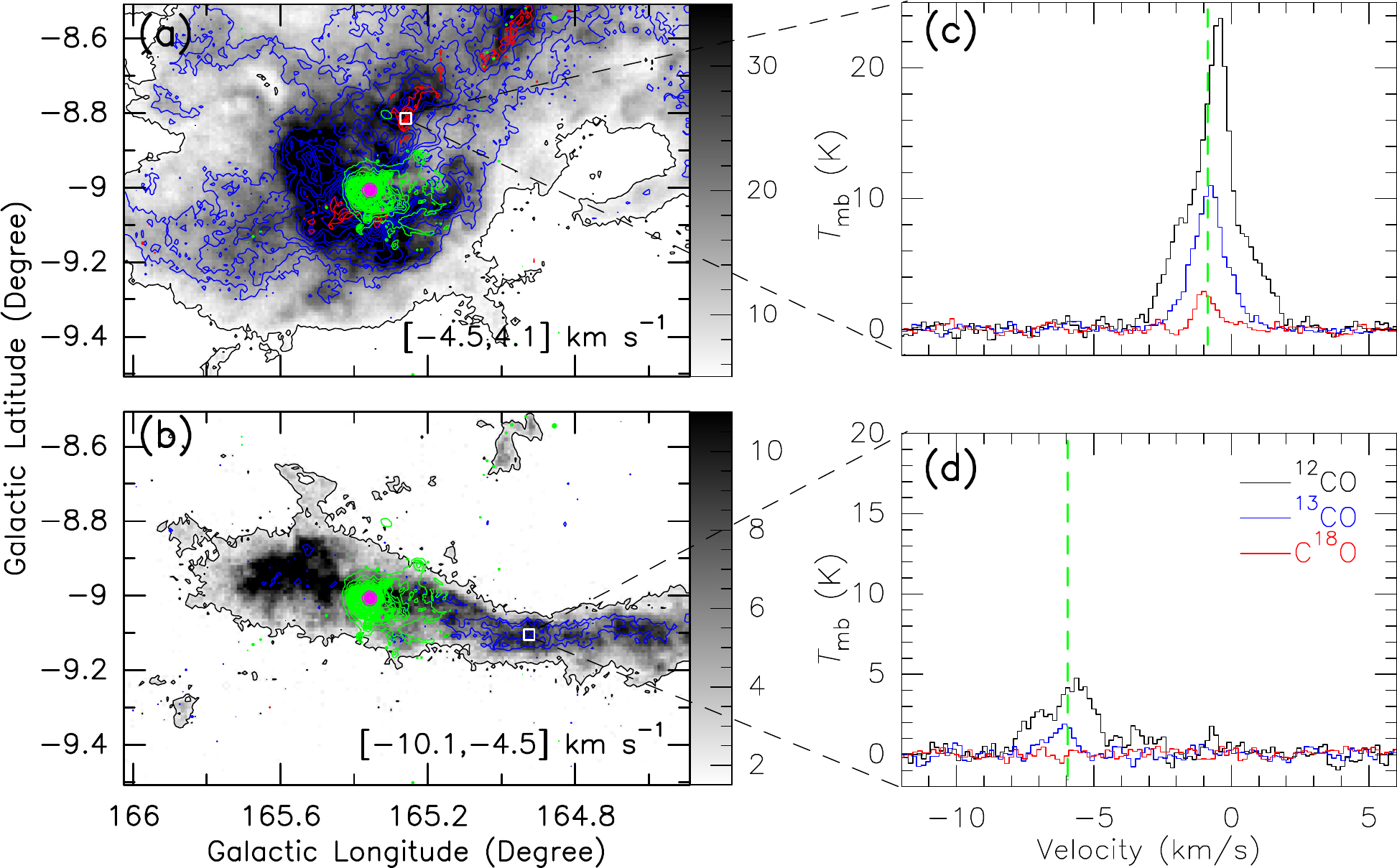}
\vspace{-1mm}
\caption{(a) and (b) panels: 1.4 GHz radio continuum contours (pink), WISE 12 $\mu$m contours (green), $^{13}$CO $J$=1-0 integrated-intensity contours (blue), and C$^{18}$O $J$=1-0 contours (red) superimposed on the integrated intensity $^{12}$CO $J$=1-0 maps (grey). The integrated velocity ranges are indicated in the lower right corner of each map. The right grey bar is units of K km s$^{-1}$. The contours begin at 5$\sigma$ in each emission. (c) and (d) panels: The typical spectra of the positions are indicated by the white squares in Figures \ref{Fig:Integrated intensity}(a) and \ref{Fig:Integrated intensity}(b).}
\label{Fig:Integrated intensity}
\end{figure*}

H$_{2}$ column density only gives the distribution of projected 2D gas. Here, we use $^{12}$CO $J$=1-0, $^{13}$CO $J$=1-0, and C$^{18}$O $J$=1-0 lines to trace the molecular gas distribution of the two filaments.  To investigate the gas structure of the minor filament, we made a position-velocity (PV)  diagram (Figure \ref{Fig:PV}) of the $^{12}$CO $J$=1-0 emission along the black dashed line and through the position of the  LkH$\alpha$ star in Figure \ref{Fig:SH2-222-RGB}. In Figure \ref{Fig:PV}, we see that there are two main velocity components, which range from -10.1 km s$^{-1}$ to -4.0 km s$^{-1}$, and from -4.0 km s$^{-1}$ to 4.1 km s$^{-1}$. In $^{12}$CO $J$=1-0 emission, we can clearly see bridging features in velocity between the two components, indicated by the green dashed lines in Figure \ref{Fig:PV}.   Moreover, the $^{12}$CO $J$=1-0 emission show a molecular shell and a ``V-shape'' gas structure, which are indicated by the green circle and yellow dashed lines in Figure \ref{Fig:PV}. Using the two velocity ranges, we make the integrated intensity maps of $^{12}$CO $J$=1-0, $^{13}$CO $J$=1-0, and C$^{18}$O $J$=1-0,  overlaid with the 1.4 GHz radio continuum contours (pink) and WISE 12 $\mu$m contours (green), as shown in Figures \ref{Fig:Integrated intensity}(a) and \ref{Fig:Integrated intensity}(b). In Figure \ref{Fig:Integrated intensity}(a), the velocity component of -4.0 km s$^{-1}$ to 4.1 km s$^{-1}$ displays a filamentary structure elongated from southeast to northwest, which should be belong to the main filament. For the main filament, the $^{12}$CO $J$=1-0 and $^{13}$CO $J$=1-0 emission also show a shell-like structure, just surrounding the LkH$\alpha$ 101 \HII region. From the PV diagram, we also see the molecular shell. Compared with the $^{12}$CO $J$=1-0 and $^{13}$CO $J$=1-0 emission, the C$^{18}$O $J$=1-0 emission may trace the dense part of the main filament. The C$^{18}$O $J$=1-0 emission is likely to be cut into two parts at the position of the LkH$\alpha$ 101 \HII region. In Figure \ref{Fig:Integrated intensity}(b), the velocity component of -10.1 km s$^{-1}$ to -4.0 km s$^{-1}$ shows a northeast-southwest filament, whose direction is almost perpendicular to the main filament.  Compared with Figure \ref{Fig:SH2-222-RGB}(a), the northeast-southwest filament is associated with the minor filament. We did not detect the C$^{18}$O $J$=1-0 emission in the whole minor filament, suggesting that the minor filament is not a filament like the main filament with dense gas. 

Figures \ref{Fig:Integrated intensity}(c) and \ref{Fig:Integrated intensity}(d) give
the spectra in the typical positions of the two filaments. From the C$^{18}$O $J$=1-0 spectrum, we obtain that the systemic velocity of the main filament is -0.7 km s$^{-1}$.  Since we do not detect the  C$^{18}$O $J$=1-0 emision in the minor filament, the systemic velocity of -6.0 km s$^{-1}$ for the minor filament is determined by  the $^{13}$CO $J$=1-0 emision.  The systemic velocities of the two filaments are marked in blue and red dashed lines in Figure \ref{Fig:PV}, suggesting that although the two filaments intersect each other, the intersection is characterized by two different velocity components.  The LkH$\alpha$ 101 embedded cluster with the \HII region is just located at the projected intersectional region of the two filaments in the line of sight.   

\subsection{Young Stellar Object Distribution}
Based on two young stellar object (YSO) catalogs, \citet{Lada2017} summarized a new YSO table, which contains 43 Class I YSOs, 116 Class II YSOs, and 7 Class III YSOs in the whole CMC.  Class I YSOs are protostars with  circumstellar envelopes and a timescale of the order of $\sim10^{5}$ yr, while Class II YSOs are disk-dominated objects with a $\sim10^{6}$ yr \citep{Andre94}. The Class III YSOs are the pre-main sequence stars. From the table, we found 18 Class I YSOs, 72 Class II YSOs, and 4 Class III YSOs in our observed region. Figure \ref{Fig:YSO}(a) shows the spatial distribution of these YSOs. To investigate the positional relation of YSOs with the main and minor filaments, we overlaid the selected YSOs on the $^{13}$CO $J$=1-0 emission map (green) of the main filament, and the $^{12}$CO $J$=1-0 emission map (blue) of the minor filament. Class I and Class II YSO sources in Figure \ref{Fig:YSO}(a) are found to be mainly distributed along the main filament. The main filament contains $\sim 42\%$ of Class I YSOs and  $\sim 62\%$ of Class II YSOs in the CMC while occupying only a small fraction of the total cloud area. The main filament includes most young stars, probably because it is the densest part of the CMC, and has a high gas density. In Figure \ref{Fig:YSO}(a), we  can also clearly see that compared with other region in the main filament, the projected intersectional region of the two filaments gathers more Class II YSOs with about 1 Myr.  

\section{Discussion}
\subsection{Dynamics structure}
Through the Herschel H$_{2}$ column density and CO emission maps, we found the two filaments intersect each other in the line of sight. As shown in the Figure \ref{Fig:PV}, the main and minor filaments have different systemic velocities.  It suggests that the intersection from the two filaments is characterized by two different velocity components. This scenario indicates that the two filaments is likely to be colliding with each other. We also find several pieces of evidence to support the colliding idea. In the $^{12}$CO $J$=1-0 PV map, the bridging features connect the two filaments in velocity, which indicates that the two filaments are interacting \citep[e.g.,][]{Haworth2015,Torii2017,Fukui2018}.  In addition, the main filament shows a ``V-shape'' gas structure in the PV diagram, which is similar to that shown in Figure 14 of \citet{Fukui2018}. \citet{Fukui2018} suggested that the V-shaped protrusion including the bridge in a  PV diagram provide a characteristic feature of collision, especially for that one of the colliding clouds is smaller than the other. We also know that the size of the minor filament is significantly smaller than that of the main filament. For the main filament, the $^{12}$CO $J$=1-0 and $^{13}$CO $J$=1-0 emission present a shell-like structure, which surrounds a PAH bubble created by the  LkH$\alpha$ 101 \HII region. The shell-like structure detected indicates that the expanding PAH bubble created by the LkH$\alpha$ 101 \HII region is interacting with the main filament. However, from Figure \ref{Fig:PV}, we see that the LkH$\alpha$ 101 \HII region is still confined to a molecular shell, which is shown in a green circle. Furthermore, the two bridging features in Figure \ref{Fig:PV} are not connected to the LkH$\alpha$ 101 \HII region, indicating that the bridging and V-shape features are not created by the feedback of the LkH$\alpha$ 101 \HII region. Hence, we further conclude that the main filament is colliding with the minor filament. 

\begin{figure}
\centering
\includegraphics[width = 0.36 \textwidth]{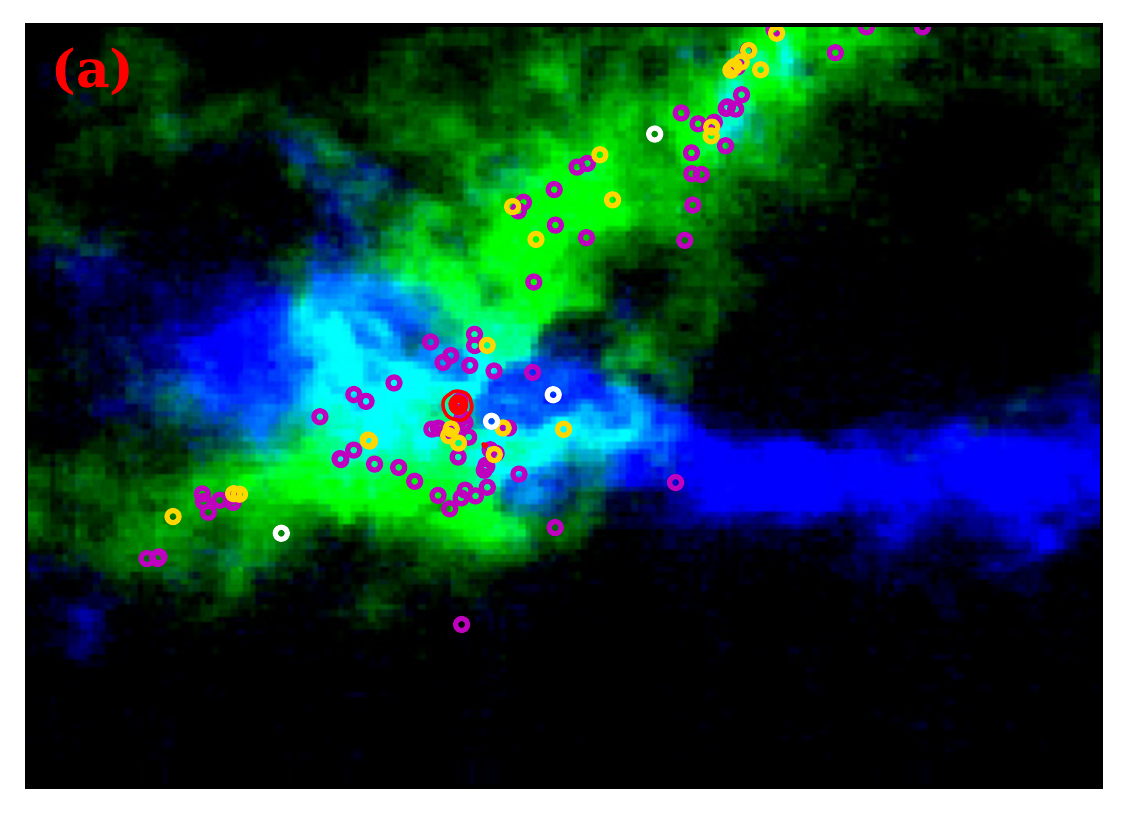}
\includegraphics[width = 0.36 \textwidth]{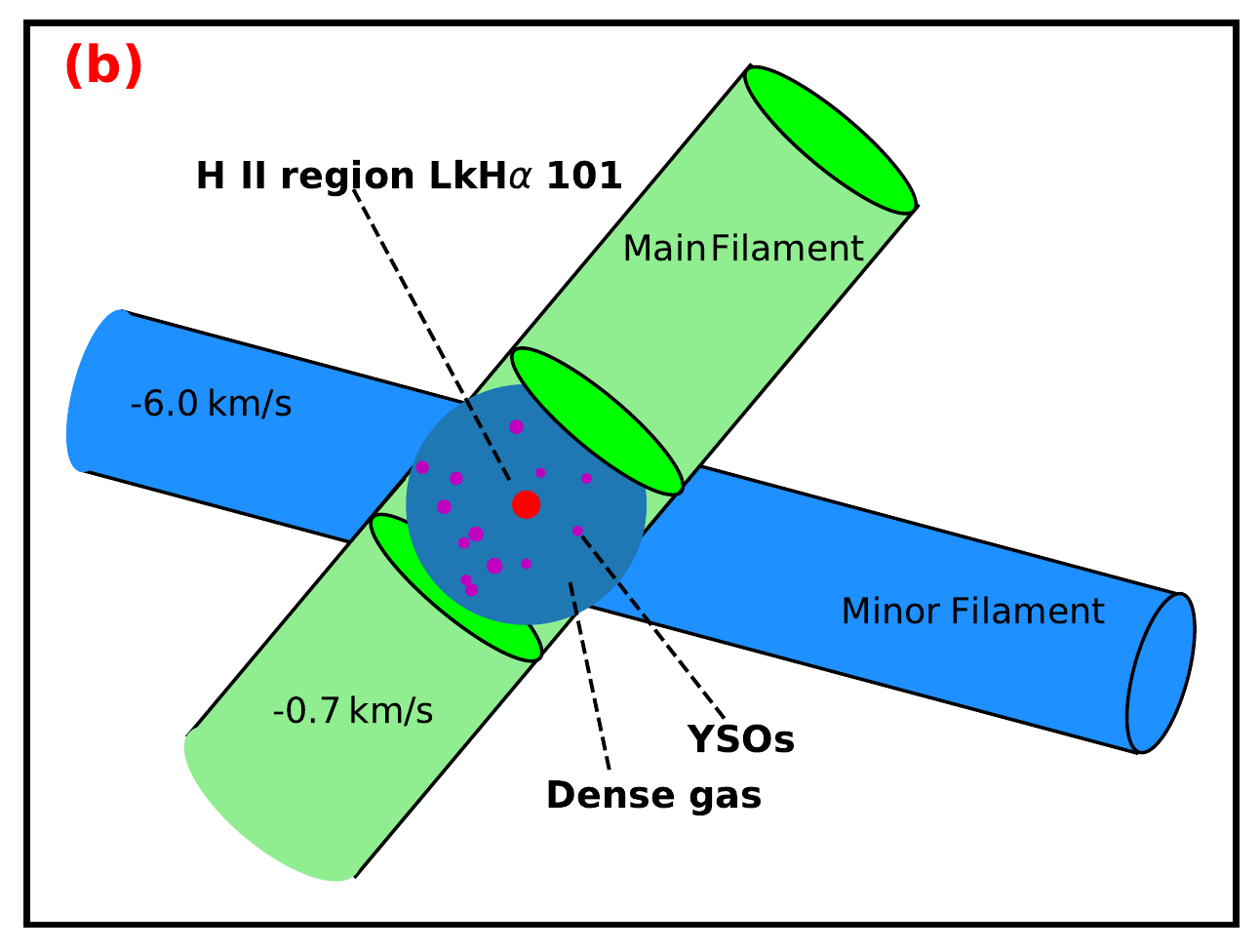}
\caption{(a) panel: 1.4 GHz radio continuum contours (red colour) overlaid on the $^{13}$CO $J$=1-0 emission map (green) of the main filament, and $^{12}$CO $J$=1-0 emission map (blue) of the minor filament. The pink circles indicate the positions of {\bf Class} II YSOs, the yellow and white circles for {\bf Class} I YSOs and {\bf Class} III YSOs, respectively. (b) panel: sketched diagram of the main and minor filaments.}
\label{Fig:YSO}
\end{figure}

\subsection{The LkHα 101 embedded cluster formation}
In the CMC or main filament, the LkH$\alpha$ 101 embedded cluster with an \HII region and several B0.5 stars is the first and only one significant cluster \citep{Barsony1990,Dzib2018}. We see that the LkH$\alpha$ 101 embedded cluster is just located at the projected intersectional position of the main filamnet with the minor filament. The morphology can be alternatively explained by the filament-hub accretion from the surrounding filament material, which was believed to play an important role in the formation of the young embedded cluster \citep{Myers2009,Kirk2013}, however, this does not seem to be the case for the LkH$\alpha$ 101 embedded cluster formation in the main filament. We can adopt the RRL velocity (0.5$\pm$0.1 $\kms$) of the LkH$\alpha$ 101 \HII region as the system velocity of the LkH$\alpha$ 101 embedded cluster,  which is not located between -10.1 km s$^{-1}$ and -4.0 km s$^{-1}$ for velocity component of the minor filament. Hence, the LkH$\alpha$ 101 embedded cluster has no direct contact with the minor filament, and it is impossible for the minor filament to provide material for the embedded cluster formation.  \citet{Li2014} proposed that the formation of the embedded cluster was caused by the merging of the two sub-filaments in the main filament. It is very similar to the hub-system accretion model. Furthermore, if there is an accretion flow in the minor filament, such accreting flows will display self-absorption in optically thick emission \citep{Lee2013,Kirk2013}, which is not found in our $^{12}$CO $J$=1-0 and $^{13}$CO $J$=1-0 data in the main and minor filament, as shown in the Figures \ref{Fig:Integrated intensity} (c) and (d), respectively. 

In addition, the cross-like structure combined by the main and minor filaments seem to be similar to L1188. \citet{Gong2017} suggested that cloud-cloud collision happened in the L1188, and triggered star formation in it. From Section 4.1, we know that the cloud-cloud collision exactly happened in the main filament. The age of the LkH$\alpha$ 101 embedded cluster would be 1-2 Myr \citep{Herbig2004}, suggesting that the LkH$\alpha$ 101 embedded cluster appears to be an individual embedded cluster. This inquires that a significant amount of dense gas need to be rapidly gathered in the position of the main filament where the LkH$\alpha$ 101 embedded cluster was born. Recent magnetohydrodynamic (MHD) numerical simulations on cloud-cloud collision show that after the collision the local density can rapidly increases from 300-1000 cm$^{-3}$ to 10$^{4}$ cm$^{-3}$ \citep{Inoue2013}, suggesting that the cloud-cloud collision can provide help for an embedded cluster formation.  Hence, we conclude that the collision between the main and minor filaments may play a certain role in the LkH$\alpha$ 101 embedded cluster formation.

The system velocity of the LkH$\alpha$ 101 embedded cluster is just located between -4.0 and 4.1 km s$^{-1}$ for the velocity component of the main filament, indicating that the LkH$\alpha$ 101 embedded cluster formed in the main filament. The formation circumstances of the LkH$\alpha$ 101 embedded cluster in the CMC is very similar to that of super star cluster RCW 38 \citep{Fukui2016}. The triggered clusters form in the main component, not in the interaction region of two molecular clouds. Figure \ref{Fig:YSO}b shows a summary for the formation scenario of the LkH$\alpha$ 101 embedded cluster.  A super star cluster formation is triggered where two clouds collide at a velocity separation of 10-30 km s$^{-1}$ \citep{Fukui2016}.  For the main and minor filaments, the velocity separation is 5.3 km s$^{-1}$. Therefore, here cloud-cloud collision triggered the formation of the LkH$\alpha$ 101 embedded cluster with B stars, not a super star cluster. 

In Figure \ref{Fig:YSO}(a), there are some YSOs distributed outside the intersectional region in the main filament. The filament can be fragmented to form prestellar cores and protostars \citep{Andre2010}. Therefore, the evolution and fragmentation of the main filament may also plays an important role in some YSOs formation of the  LkH$\alpha$ 101 embedded cluster. However, these YSOs don't seem to form a real cluster, suggesting that except for the intersectional region, the other regions seem to have no conditions to produce an embedded cluster in the main filament. Hence, we suggest that the cloud-cloud collision together with the fragmentation of the main filament help the LkH$\alpha$ 101 embedded cluster formation in the CMC. Although the present observations can explain the formation of the LkH$\alpha$ 101 embedded cluster, the more observations at higher spatial resolution are needed to resolve the detailed kinematics in this region.

\section{CONCLUSIONS}
In this Letter, we present a multi-wavelength observation towards LkH$\alpha$ 101 embedded cluster and its adjacent region. These observations have revealed that the embedded cluster are just located at the  projected intersectional region of the main and minor filaments. The main filament is the highest-density part of the CMC, while the minor filament is a new identified filament with a low-density gas emission. In the $^{12}$CO $J$=1-0 PV map, the intersection shows the bridging features connecting the two filaments in velocity, and we also identify a V-shape gas structure. These agree with the scenario that the two filaments are colliding with each other. From the FAST observation, we obtained the RRL velocity of 0.5$\pm$0.1 $\kms$ for the LkH$\alpha$ 101  H II region, which is just located between -4.5 to 4.1 $\kms$ for the velocity component of the main filament.  We suggest that the collision between the two filaments may play a certain role the LkH$\alpha$ 101 embedded cluster formation in the CMC.  In addition, there are also some YSOs distributed outside the intersectional region, this indicates that the evolution and fragmentation of the main filament may also play an important role in the YSOs formation of the cluster.
  
\acknowledgments We thank the referee for insightful comments that improved the clarity of this manuscript. This work was supported by the Youth Innovation Promotion Association of CAS, the National Natural Science Foundation of China (Grant Nos. 11873019 and 11933011), and also supported by the Open Project Program of the Key Laboratory of FAST, NAOC, Chinese Academy of Sciences.

\end{document}